\documentclass[aps,prl,twocolumn,superscriptaddress,groupedaddress]{revtex4}
\usepackage{amssymb}
\usepackage{amsmath,bm}
\usepackage{graphicx,color}
\usepackage{bbm}
\newcommand{\B}[1]{{\bm{#1}}}

\begin{document}

\title{Oscillatory Instabilities in Frictional Granular Matter}

\author{Joyjit Chattoraj} 
\affiliation{School of Physical and Mathematical Sciences, Nanyang Technological University, Singapore}
\author{Oleg Gendelman} 
\affiliation{Faculty of Mechanical Engineering, Technion, Haifa 32000, Israel}
\author{Massimo Pica Ciamarra}
\email{massimo@ntu.edu.sg}
\affiliation{School of Physical and Mathematical Sciences, Nanyang Technological University, Singapore}
\affiliation{CNR--SPIN, Dipartimento di Scienze Fisiche,
Universit\`a di Napoli Federico II, I-80126, Napoli, Italy}
\author{Itamar Procaccia}
\affiliation{Department of Chemical Physics, the Weizmann Institute of Science, Rehovot 76100, Israel}

\begin{abstract}
Frictional granular matter is shown to be fundamentally different in its plastic responses to
external strains from generic glasses and amorphous solids without friction. While regular glasses exhibit plastic instabilities due to a vanishing of a real eigenvalue of the Hessian matrix, frictional granular materials can exhibit a previously unnoticed additional mechanism for instabilities, i.e. the appearance of a pair of complex eigenvalues leading to oscillatory exponential growth of perturbations which are tamed by dynamical nonlinearities. This fundamental difference appears crucial for the understanding of plasticity and failure in frictional granular materials. The possible relevance to earthquake physics is discussed.
\end{abstract}

\maketitle

It is often stressed that the mechanical properties of frictional granular matter and of
glassy amorphous solids share many similarities \cite{98LN,01OLLN,05Wya,11BB,Ciamarra2011},
although the effective forces in frictional solids are not derivable from a Hamiltonian.
Here we show that the lack of a Hamiltonian description is
responsible for previously unreported oscillatory instabilities
in frictional granular matter.
These oscillatory instabilities furnish a micromechanical mechanism for a giant
amplification of small perturbations that can lead eventually to major events of mechanical failure.
We will demonstrate this physics  in the context of amorphous assemblies of frictional disks, but will
make the point that the mechanism discussed here is generic for systems with friction.
To motivate the new ideas recall that the understanding of plastic instabilities, shear banding and mechanical failure in athermal amorphous solids with an underlying Hamiltonian description had progressed significantly in the last twenty years. Beginning with the seminal papers of Malandro and Lacks \cite{98ML,99ML} it became clear that an object that controls the mechanical responses of athermal glasses is the Hessian matrix. In an athermal (T=0) system of $N$ particles at positions $(\B r_1, \B r_2\cdots \B r_N)$
we define the Hamiltonian $U(\B r_1, \B r_2,\cdots \B r_N)$. The Hessian matrix is
\begin{equation}
H^{\alpha\beta}_{ij} \equiv \frac{\partial^2 U(\B r_1, \B r_2,\cdots \B r_N)}{\partial r^\alpha_i \partial r^\beta_j} = -\frac{\partial F^\alpha_i}{\partial r^\beta_j}\ .
\end{equation}
Here $\B F_i$ is the total force on the $i$th particle, and in systems with binary interactions we can write
$\B F_i \equiv \sum_j \B F_{ij}$ with the sum running on all the particles $j$ interacting with particle $i$.
Being real and symmetric, the Hessian matrix has real eigenvalues which are all positive as long as the material is mechanically stable.
Under strain, the system may display a saddle node
bifurcation in which an eigenvalue goes to zero, accompanied by a localization of an
eigenfunction, signalling a plastic instability that is accompanied by a drop in stress and
energy \cite{04ML}.
Significant amount of work was dedicated to understanding the density of states
of the Hessian matrix which differs in amorphous solids from the classical Debye density
of purely elastic materials \cite{05Wya,10KLP,Mizuno2017}.
The well known ``Boson peak" was explained by the prevalence
of ``plastic modes" that can go unstable and do not exist in pure elastic systems. The
system size dependence of the eigenvalues of the Hessian \cite{10KLP}, their role in determining the mechanical characteristics like the elastic moduli \cite{11HKLP}, the failure of nonlinear elasticity in such materials \cite{11HKLP,16PRSS,17DIPS},
the relevance to shear banding and mechanical failure \cite{12DHP,13DHP,13DGMPS}, all underline the importance of this approach to the theory of amorphous solids.

Alas, this useful approach appears to be irretrievably lost when we consider the available models for {\em frictional} granular media with both normal and tangential forces at every contact of two granules. The reason is two-fold. First, the tangential forces $\B F_{ij}^{(t)}$ (see below for details),
are not analytic because of the Coulomb constraint, $\left| \B F_{ij}^{(t)} \right| \le \mu \left| \B F_{ij}^{(n)}\right|$, bounding the magnitude of the tangential force by the normal force $\B F_{ij}^{(n)}$ multiplied by $\mu$ which is the friction coefficient.
Secondly, and most importantly, model forces in frictional granular systems are not derivable from a
Hamiltonian. In the most popular models, like the Hertz-Mindlin model \cite{49Min}, the inter-particle
forces are derived by coarse graining the highly complex microscopic mechanics of compressed granules.
As the resulting model forces cannot be derived from a Hamiltonian function, they are not energy conserving. We stress that this occurs also in the absence of viscous damping and before the Coulomb limit is reached.

To describe the failure of a granular systems as a dynamical instability we follow
a two step approach. The first (maybe trivial looking) step that we propose here is to smooth out the approach to the Coulomb limit to allow differentiating the tangential force, and see Eq.~(\ref{Ft}) below. In the second
step we consider frictional disks for which the coordinates now include the positions $\B r_i$ of the centers
of mass and the angles $\theta_i$ of each disk.  The Newton equations of motion are written as 
\begin{equation}
m_{i}\frac{d^2 \B q_{i}}{dt^2}={\B F}_i(\B q_1,\B q_2,\cdots,\B q_N)
\label{Newton}
\end{equation}
where $\B q_i\equiv \{\B r_i,\theta_i\}\equiv \{r_i^x, r_i^y,\theta_i\}$ and 
$m_i$ are masses or moments of inertia as is appropriate.
It is important to stress that the forces in Eq.~(\ref{Newton}) depend only on the generalized coordinates $\B q_j$, i.e. first derivatives do not appear. The stability of equilibria of Eq.~(\ref{Newton}) is then determined by an operator obtained from the derivatives of the force $\B F_i$ on each particle with respect to the coordinates. In other words
\begin{equation}
\B J_{ij} \equiv -\frac{\partial \B F_i}{\partial \B q_j}.
\end{equation}
The analogy between the operator $\B J$ and the Hessian matrix is apparent.
But there is a huge difference whose consequences are explored below.
$\B J$ is not a symmetric operator. Accordingly, it can have real eigenvalues
as the Hessian, but it can also display a number of eigenvalues as complex conjugate pairs.
When a pair complex eigenvalues, $\lambda_{1,2}=\lambda_r\pm i\lambda_i$, gets born, a novel instability mechanism develops. Indeed, these eigenvalues
correspond to FOUR solutions $e^{i\omega t}$ to the linearized equation
of motion with
\begin{equation}
i\omega_{1,2} = \omega_i \pm i\omega_r\ , \quad i\omega_{3,4}=-\omega_i \pm i\omega_r\ .
\label{four},
\end{equation}
with $\omega_r \pm i\omega_i = \sqrt{\lambda_r \pm i\lambda_i}$.
The first pair in~(\ref{four}) will induce an oscillatory motion with an exponential growth of any deviation $\B q(0)$ from a state of mechanical equilibrium,
\begin{equation}
 \B q(t) = \B q(0) e^{\omega_i t} \sin(\omega_r t).
 \label{growth}
\end{equation}
The second pair represents an exponentially decaying oscillatory solution.
We stress that this bifurcation is not a regular Hopf bifurcation.
It needs at least four degrees of freedom (four first order or two second order differential equations). 
This is a somewhat unusual bifurcation that is appearing here due to the symmetry of Eqs.~(\ref{Newton}) that is
a consequence of the absence of first derivatives. We also comment again that such a bifurcation is impossible in frictionless amorphous solids with a microscopic Hamiltonian.

To validate this theoretical scenario and explore its consequences  we focus
on a binary assembly of $N$ frictional disks of mass $m$ in a box of size $L^2$, half of which with radius $\sigma_1=0.5$ and the other half with $\sigma_2=0.7$. Under external stress they interact with binary interactions; the normal
force is determined by the overlap $\delta_{ij} \equiv \sigma_i+\sigma_j-r_{ij}$ where
$\B r_{ij}\equiv \B r_i- \B r_j$. The normal force is Hertzian,
\begin{equation}
\B F_{ij}^{(n)} = k_n \delta_{ij}^{3/2}\hat r_{ij} \ , \quad \hat r_{ij} \equiv \B r_{ij}/r_{ij} \ .
\label{Fn}
\end{equation}
The tangential force is determined by the tangential displacement $\B t_{ij}$,
the integral of the velocity at the contact point over the duration
of the contact, rotated so as to enforce $\B t_{ij} \cdot \hat r_{ij} = 0$ at all times.
This is quite standard~\cite{01SEGHLP}.
We deviate from the standard in the definition of the tangential force,
that we assume to be
\begin{eqnarray}
&&\B F_{ij}^{(t)} = -k_t\delta_{ij}^{1/2}\left[1+\frac{t_{ij}}{t^*_{ij}} -\left(\frac{t_{ij}}{t^*_{ij}}\right)^2\right]t_{ij} \hat t_{ij} \ , \nonumber\\
&&t^*_{ij} \equiv \mu \frac{k_n}{k_t} \delta_{ij}
\label{Ft}
\end{eqnarray}
with $k_t=2k_n/7$~\cite{01SEGHLP}.
The derivative of the force with respect to $t_{ij}$ vanishes smoothly at $t_{ij}=t^*_{ij}$,
and the Coulomb law is fulfilled.
In the following, we use as units of mass, length and time $m$, $2\sigma_1$
and $\sqrt{m(2\sigma_1)^{-1/2}k_n^{-1}}$, respectively.
We also fix the friction coefficient to a high value, $\mu = 10$, to emphasise that the existence of a Coulomb threshold is no responsible for the reported phenomenology, but we stress here that we have found analogous results for values of $\mu < 1$.

We demonstrate the new type of instability considering a system with $N=500$.
We prepare a mechanically equilibrated amorphous system with packing fraction 0.93 in a periodic 2-dimensional box. Upon straining we can choose to run two types of algorithms. The first is denoted Newtonian and is simply a solution of the Newton equations of motion with the given forces Eqs.~(\ref{Fn}) and (\ref{Ft}) without damping.
The second algorithm is called ``overdamped" and is solving the same equations of motion but with a damping force that is proportional to the velocities of the disks with
a coefficient of proportionality $\eta_v=m\eta_0$.
We fix $\eta_0 = 10$.
The damping timescale $\eta_v^{-1}$ is thus of the order of the time that
sounds needs to travel one particle diameter~\cite{Zhang2005}, making this
dynamics overdamped.
With damping, even in the presence of complex eigenvalues the oscillatory instability
is suppressed by the damped dynamics.
The numerical solution of the equation of motion is carried
out with LAMMPS~\cite{Plimpton1995}. 

\begin{figure}
    \includegraphics[width=0.30\textwidth]{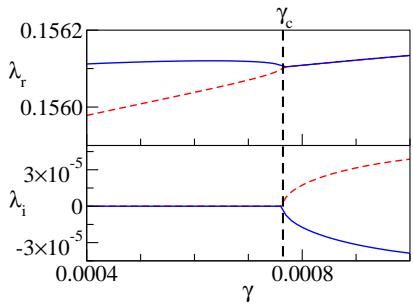}
    \caption{Upon increasing the strain $\gamma$ two modes with real eigenvalues $\lambda$ coalesce at $\gamma_{c}$ (dashed vertical lines), and a pair of complex conjugate modes gets born. The upper and the lower panels show the evolution of the real and of the imaginary components of these modes.}
    \label{bifurcation}
\end{figure}

An athermal quasi static (AQS) shear protocol is now devised as follows: starting from the initial stable configuration the system is sheared along the horizontal direction ($x$) by the amount $\delta\gamma$, varied in the range $10^{-4}$ to  $10^{-8}$ depending on the precision needed for the identification of the instability. Thus, each particle experiences an affine shift along $x$ depending on their vertical coordinates $r^y_i$, i.e.~$\delta r_i^x = \delta\gamma r^y_i$. Next we run the overdamped dynamics to bring the system back to mechanical equilibrium where the net force on each particle is less than $10^{-8}$.
After every such step we diagonalize the matrix $\B J$ to find its eigenvalues.
At some value of $\gamma$ we find
for the first time the birth of conjugate pair of complex eigenvalues as seen in Fig.~\ref{bifurcation}.
If we continue to increase the strain using the same protocol, we see the emergence of other complex pairs at the expense of real eigenvalues.

\begin{figure}
\vskip 1.4 cm
    \includegraphics[width=0.45\textwidth]{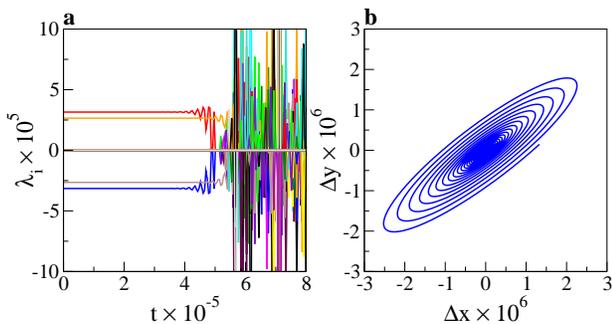}
      \caption{
      {\bf a}: Time dependence of the imaginary component of all 1500 eigenvalues of the system, during a Newtonian simulation.
      {\bf b}: typical spiral trajectory of a particle in the linear response regime.}
    \label{Newtonian}
\end{figure}
In real granular systems, the dynamics is not overdamped. To explore how the system responds
to the bifurcation we therefore run the Newtonian dynamics. As an example we do it here starting from a
configuration with two complex-conjugate eigenpairs.
The dominant eigenpair, which is the one with the largest growth rate $\omega_i$, has
$\omega_r = 0.395122$, $\omega_i =3.99\times 10^{-5}$.
During the Newtonian dynamics, we evaluate the operator ${\B J}$ and its eigenvalues.
We find that all the eigenvalues remain invariant for a long stretch of time, as illustrated in Fig.~\ref{Newtonian}a, until a major instability takes place.
An insight on the expected particle motion is obtained considering real 
matrices admit a real decomposition of the kind ${\B J} = {\B C}{\B D}{\B C}^{-1}$.
If ${\B J}$ is symmetric, then $\B D$ is the diagonal matrix containing the eigenvalues.
If ${\B J}$ is not symmetric, then $\B D$ is block diagonal.
The blocks are $1\times1$
blocks containing the real eigenvalues, or rotation-scaling blocks
$|\lambda| \B R(\theta)$ with $\B R$ $2\times 2$ rotation matrix, one block for each complex eigenvalue pair $|\lambda|e^{\pm i\theta}$. This clarifies that the complex eigenvalues, i.e.
the rotation-scaling blocks, induce a spiral motion. The investigation of a typical
particle trajectory during the development of the instability confirms this expectation,
as we illustrate Fig.~\ref{Newtonian}b. See the Supplemenray Material for an animation of the emerging motion.

Next we consider the mean-square displacement $M(t)$ as a function of time.
Denoting $\Delta r^x_i(t) \equiv r^x_i(t) -r^x_i(t=0)$ etc. we define
\begin{equation}
M(t)\equiv \frac{1}{N} \sum_i^N [(\Delta r^x_i(t))^2 + (\Delta r^y_i(t))^2 + \sigma_i^2 (\Delta \theta_i(t))^2] \ ,
 \label{defMSD}
 \end{equation}
 which according to Eq.~\ref{growth} should behave as
 $M(t) \propto  e^{2 \omega_i t} \sin^2(\omega_r t)$.
\begin{figure}
    \includegraphics[width=0.30\textwidth]{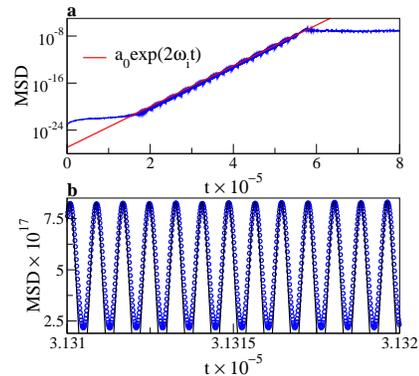}
      \caption{
      {\bf a}: The numerically computed mean-square displacement as a function of time. The red line is the predicted exponential growth from the linear instability, $a_0 e^{2\omega_it}$, with $a_0$ being fitted.
      {\bf b}: a blow up of the growth of the mean-square displacement. The black line is
      the exponential oscillatory instability prediction, $a_0 e^{2 \omega_i t} [\sin(\omega_r t+\psi)]^2$, with $\psi$ fitted.}
    \label{MSD}
\end{figure}
Indeed, we see in Fig.~\ref{MSD} that $M(t)$ shoots up in time about sixteen orders of magnitude with exponential rate and oscillatory form precisely as predicted by the linear instability. We have also checked that the rotational contribution to $M(t)$ is negligible.
We notice that the instability dominates the response after a short transient; this is consistent with the fact
that the first modes contributing to the mean square displacement are high frequency stable modes.

Finally, we focus on the virial component of the shear stress $\sigma_{xy} = -\frac{1}{L^2}\sum_{i\neq j}^{N}r^x_{ij}F_{ij}^y$.
During the development of the instability, the stress change
is predicted to evolve as $\sigma_{xy}(t) -\sigma_{xy}(0) \propto e^{\omega_i t} \sin(\omega_r t+\psi)$.
\begin{figure}
  \includegraphics[width=0.30\textwidth]{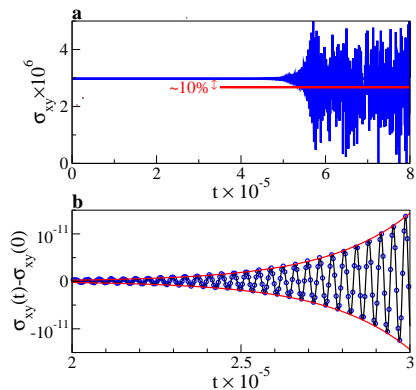}
  \caption{
    {\bf a}: the evolution of shear stress (virial contribution) $\sigma_{xy}$ during Newtonian dynamics. The instability occuring at $t\approx 5 \times 10^5$ results in a significant drop   of the average stress.
    {\bf b}: blow up of the stress change during the development of the instability. The black line is the theoretical prediction, $\Delta\sigma e^{\omega_it}\sin{\omega_rt}$, with a fitted $\Delta \sigma$; the red lines mark the envelope $\pm\Delta\sigma e^{\omega_it}$.}
  \label{stress}
\end{figure}
Fig.~\ref{stress}a shows that the stress follows the predicted linear instability with its exponential growth and oscillations until the perturbation self-amplifies enough
to induce a major plastic instability in which the system undergoes a micro earthquake and loses $\simeq 10\%$ of the stress.

Taken together, Figs~\ref{Newtonian}a, \ref{MSD}a and \ref{stress}a indicate that the predictability
of the evolution under the effect of
the oscillatory exponential instability terminates at a time $t \approx 5\times 10^5$.
Around that time the perturbation amplified enough for the system to switch on a non-linear response characterized by the coexistence of a number of unstable modes, saturating the mean square displacement,
and causing large stress fluctuations which eventually result in a significant stress drop, cf. Fig.~\ref{stress}.

At this point is is important to stress that the existence of the oscillatory instability is not 
limited to the particular choice of forces Eqs.~(\ref{Fn}) and (\ref{Ft}). Any reasonable coarse grained theory of tangential forces must
take into account the fact that compressed granules will create a larger area of contact. Accordingly,
it is expected that the tangential force will be a function not only of the $\theta_i$ coordinates but also
of the positional coordinates $\B r_i$. Consequently, in general the forces would not be derivable
from a Hamiltonian, and the corresponding operator $\B J$ will not be symmetric. There is therefore a
generic possibility to find complex eigenpairs in this operator in any reasonable coarse-grained theory of
frictional matter.

Having this genericity in mind, we would like to cautiously 
speculate about the relevance of the findings reported above to
the physics of earthquakes. We of course do not propose that the system studied above of frictional disks includes all the rich physics of the earth and its faults. Nevertheless it is tempting to consider one of the most striking observation in earthquake physics which is known as ``remote
triggering"~\cite{06FB,00BKK,14BE}: an earthquake could
trigger a subsequent
earthquake on a different fault, even if located far away. It is clear that
faults can `communicate'
via seismic waves propagating through the earth crust. Specifically, distant
faults can only communicate via
long wavelength seismic waves, as short wavelengths are quickly damped as they
propagate.
However seismic waves with long wavelengths act as small perturbations, as they
have a small frequency and hence a small energy density, so that it is not clear how
they could be able to induce the failure of a fault.
The most popular approach to rationalize this observation within
the geophysical community, is the acoustic fluidization~\cite{96Mel,Giacco2015}
mechanism. This mechanism was invoked to rationalize remote triggering, suggesting
that long wavelengths impacting on a fault trigger short wavelengths within the
fault, and that these act by reducing the confining pressure and promoting failure.
However, a detailed micromechanical investigation
of this process is lacking. We would like to propose that the mechanism discussed in
this Letter might be relevant for the discussion of remote triggering.  Admittedly, our model system is too simple to resemble a geological
fault. We propose however that the mechanism that we highlight here is generic in mechanical systems
that are frictional and their dynamics is not derivable from a Hamiltonian. The crucial observation
is that we have a clear mechanism for the self-amplification of small perturbations, making
it quite worthwhile to study this mechanism also in the context of fault dynamics and in
other context of frictional granular matter.

\acknowledgments
This work had been supported in part by the ISF-Singapore exchange program and the by
the US-Israel Binational Science Foundation. We thank Jacques Zylberg and Yoav Pollack
for useful discussions and exchanges at the early stages of this project.
JC and MPC acknowledge NSCC Singapore for granting the computational facility under project 12000621.

\bibliography{ALL}
\newpage

\onecolumngrid

\section{Supplementary Material}
The operator ${\B J}$ involves controlling the time evolution of the system involves
the derivative of the forces and of the torques acting on the particles with respect
to the degree of freedom. In this supplementary notes, we first describe the interaction
forces (Sec.~\ref{sec:forces}).The tangential force depends on a tangential displacement,
whose dependence on the degree of freedom is detailed in (Sec.~\ref{sec:tangential}).
Finally, we consider the different derivatives forces as needed to evaluate the operator ${\B J}$
in Sec.~\ref{sec:J}.

\section{Interaction force~\label{sec:forces}}
In our simulation, a pair of granular particles interacts when they overlap. The overlap distance $\delta_{ij}$ is measured as
\begin{equation}
  \delta_{ij}=\sigma_i+\sigma_j-r_{ij},
  \label{Eq:delta}
\end{equation}
where $r_{ij}$ is the center-to-center distance of a pair-$i$ and $j$, and $\sigma_i$ is the radius of particle-$i$. The pair vector ${\B r}_{ij}$ is defined as
\begin{equation}
  {\B r}_{ij}=\B r_i - \B r_j.
  \label{Eq:rij}
\end{equation}
The pair-interaction force ${\B F}_{ij}$ has two contributions. ${\B F}^n_{ij}$ is the force acting along the normal direction of the pair $\hat{r}_{ij}$, and ${\B F}^t_{ij}$ is the force acting along the tangential direction of the pair ${\hat t}_{ij}$. 
The normal force is Hertzian:
\begin{equation}
  {\B F}^n_{ij}= k_n\delta^{3/2}_{ij} {\hat r}_{ij},
  \label{Eq:Fn}
\end{equation}
where $k_n$ is the force constant with dimension: Force per length${}^{3/2}$.
The tangential force ${\B F}^t_{ij}$ is a function of both the overlap distance $\delta_{ij}$ and the tangential displacement ${\B t}_{ij}$.
We have modified the standard expression for ${\B F}^t_{ij}$ and included a few higher order terms of $t_{ij}$ (i.e., $|{\B t}_{ij}|$) such that the derivative of the force function $F^t_{ij}$ with respect to tangential distance $t_{ij}$ becomes continuous and it goes to zero smoothly.
We use the following form:
\begin{equation}
  \label{Eq:Fs}
  \begin{split}
    {\B F}^t_{ij} & = -k_t\delta^{1/2}_{ij}\left[1 + \frac{t_{ij}}{t^*_{ij}} - \left(\frac{t_{ij}}{t^*_{ij}}\right)^2 \right] t_{ij}{\hat t}_{ij}\\
    & = -k_t \delta^{1/2}_{ij} t^*_{ij}{\hat t}_{ij}, \ \ \ \textrm{if}\ \ k_t\delta^{1/2}_{ij}t_{ij} > \mu |{\B F}^n_{ij}|,
  \end{split}
\end{equation}
where $k_t$ is the tangential force constant. Its dimension is force per length${}^{3/2}$. $t^*_{ij}$ is the threshold tangential distance:
\begin{equation}
t^*_{ij} = \mu \frac{k_n}{k_t} \delta_{ij},
\label{Eq:s*}
\end{equation}
where $\mu$ is the friction coefficient, a scalar quantity, which essentially determines the maximum strength of the tangential force with respect to the normal force at a fixed overlap $\delta_{ij}$. The derivative of $F^t_{ij}$ with respect to $t_{ij}$ vanishes at $t^*_{ij}$, as it turns out
\begin{equation}
 \label{Eq:dFsds}
  \begin{split}
    \frac{\partial F^t_{ij}}{\partial t_{ij}} & = k_t\delta^{1/2}_{ij}\left[1 + 2\frac{t_{ij}}{t^*_{ij}} - 3\left(\frac{t_{ij}}{t^*_{ij}}\right)^2 \right] \\
    & = 0, \ \ \ \textrm{if}\ \ k_t \delta^{1/2}_{ij}t_{ij} > \mu |{\B F}^n_{ij}|.
  \end{split}
\end{equation}

{\bf We stress here that the above forces imply a non Hamiltonian dynamics.} That is, there is not
a function $U(\delta,t)$ such that $F_n = -\frac{\partial U}{\partial \delta}$ and 
$F_t = -\frac{\partial U}{\partial t}$.


\section{Tangential displacement:\label{sec:tangential}}
The computation of the operator ${\B J}$ involves derivatives of the tangential force with respect
to the degrees of freedom, e.g. $\frac{\partial {F^t_{ij}}^\beta}{\partial r^\alpha_i}$. Since the tangential force is expressed in terms of the tangential displacement $\B t$, using the chain rule we will express these derivatives in terms of $\frac{\partial {t_{ij}}^\beta}{\partial r^\alpha_i}$. Here we evaluate these derivatives.

The derivative of tangential displacement ${\B t}_{ij}$ with respect to time $t$ is
\begin{equation}
  \frac{\mathrm d {\B t}_{ij}}{\mathrm d t} = {\B v}_{ij} - {\B v}^n_{ij} + \hat{r}_{ij}\times(\sigma_i{\B \omega}_i + \sigma_j{\B \omega}_j),
  \label{Eq:dsdt}
\end{equation}
where ${\B v}_{ij}={\B v}_i-{\B v}_j$ is the relative velocity of pair-$i$ and $j$. ${\B v}^n_{ij}$ is the projection of ${\B v}_{ij}$ along the normal direction $\hat{r}_{ij}$. ${\B v}_{ij} - {\B v}^n_{ij}$ is the tangential component of the relative velocity. 
${\B \omega}_i$ and ${\B \omega}_j$ are the angular velocity of $i$ and $j$, respectively. 
In differential form, the above equation reads:
\begin{equation}
  \mathrm d {\B t}_{ij} =  \mathrm d{\B r}_{ij} -  (\mathrm d{\B r}_{ij}\cdot\hat{r}_{ij})\hat{r}_{ij} + \hat{r}_{ij}\times(\sigma_i\mathrm d{\B \theta}_i + \sigma_j\mathrm d{\B \theta}_j),
  \label{Eq:ds}
\end{equation}
where $\mathrm d{\B \theta}_i$ is the angular displacement of $i$ which follows the relation: $\mathrm d{\B \omega}_i = \frac{\mathrm d {\B \theta}_i}{\mathrm d t}$.

Here on, we assume the two-dimensional ({$\bf 2D$}) system. 
Therefore, $\omega_i$, and so $\theta_i$, only have one component along $\hat{z}$, perpendicular to the xy plane, and $\hat{r}_{ij}\times\mathrm d{\B \theta}_i = \mathrm d \theta_i (y_{ij}\hat{x}-x_{ij}\hat{y})/r_{ij}$. This allows to write Eq.~(\ref{Eq:ds}) as
\begin{equation}
  \mathrm d t^\alpha_{ij} =  \mathrm d r^\alpha_{ij} -  (\mathrm d{\B r}_{ij}\cdot\hat{r}_{ij})\frac{r^\alpha_{ij}}{{r}_{ij}} + (-1)^\alpha(\sigma_i\mathrm d \theta_i + \sigma_j\mathrm d\theta_j)\frac{r^\beta_{ij}}{{r}_{ij}},
  \label{Eq:dsalpha}
\end{equation}
where $\alpha$ and $\beta$ can take value 0 and 1 which correspond to x and y components, respectively. Now if particle-$i$ changes its position the angular displacement remains unaffected, i.e.~$\frac{\mathrm d \theta_i}{\mathrm d r^\alpha_i}=0$. Thus, the change in tangential displacement along $\beta$ due to the change in position of particle-$i$ along $\alpha$ only contributes in translations, and it can be written as (using~(\ref{Eq:dsalpha}))
\begin{equation}
  \frac{\mathrm d t^\beta_{ij}}{\mathrm d r^\alpha_i} = \Delta_{\alpha\beta} - \frac{r^\alpha_{ij} r^\beta_{ij}}{{r}^2_{ij}},
  \label{Eq:dsbdra}
\end{equation}
where $\Delta_{\alpha\beta}$ is the Kronecker delta which is one when $\alpha=\beta$, or else zero. Similarly, a change in rotational coordinates does not modify the particles relative distance, i.e.~$\frac{\mathrm d r^\beta_{ij}}{\mathrm d \theta_i}=0$. Thus, the change in tangential displacement along $\beta$ due to the change in $\theta_i$ is (from~(\ref{Eq:dsalpha}))
\begin{equation}
  \frac{\mathrm d t^\beta_{ij}}{\mathrm d \theta_i} = (-1)^\beta \sigma_i \frac{r^\alpha_{ij}}{{r}_{ij}}.
  \label{Eq:dsbdtheta}
\end{equation}
In the above equation $\alpha$ and $\beta$ are always different. Now the magnitude of tangential distance $t_{ij}$ can be obtained from the relation $t^2_{ij} = \sum_{\alpha} {t^\alpha_{ij}}^2$. Its differential follows $\mathrm d t_{ij} =  \sum_{\alpha} \frac{t^\alpha_{ij}}{t_{ij}} \mathrm d t^\alpha_{ij}$. The derivatives of tangential distance $t_{ij}$ with respect to $r^\alpha_{i}$ and $\theta_i$ can be expressed as
\begin{eqnarray}
  \label{Eq:dsmagdra}
  \frac{\mathrm d t_{ij}}{\mathrm d r^\alpha_i} &=&  \left(\frac{t^x_{ij}}{t_{ij}}\right)\frac{\mathrm d t^x_{ij}}{\mathrm d r^\alpha_i} + \left(\frac{t^y_{ij}}{t_{ij}}\right)\frac{\mathrm d t^y_{ij}}{\mathrm d r^\alpha_i}, \\
  \frac{\mathrm d t_{ij}}{\mathrm d \theta_i} &=&  \left(\frac{t^x_{ij}}{t_{ij}}\right)\frac{\mathrm d t^x_{ij}}{\mathrm d \theta_i} + \left(\frac{t^y_{ij}}{t_{ij}}\right)\frac{\mathrm d t^y_{ij}}{\mathrm d \theta_i}.
  \label{Eq:dsmagdtheta}
\end{eqnarray}
With the help of equations~(\ref{Eq:dsbdra}) and~(\ref{Eq:dsbdtheta}) we can solve the above two differential equations. As the tangential threshold is a linear function of overlap distance $\delta_{ij}$ (see~(\ref{Eq:s*})), it also gets modified due to a change in $r^\alpha_i$ as
\begin{equation}
  \frac{\mathrm d t^*_{ij}}{\mathrm d r^\alpha_i} = - \mu\left(\frac{k_n}{k_t}\right)\frac{r^\alpha_{ij}}{r_{ij}},
  \label{Eq:ds*dra}
\end{equation}
and it is unaffected by the change in rotation, i.e.~$\frac{\mathrm d t^*_{ij}}{\mathrm d \theta_i}=0$.

\section{Evaluation of ${\B J}$\label{sec:J}}
\subsection{Derivative of tangential force}
The derivative of tangential force (equation~(\ref{Eq:Fs})) with respect to $r^\alpha_{i}$:
\begin{eqnarray}
  \frac{\partial {F^t_{ij}}^\beta}{\partial r^\alpha_i} &=& -k_t\frac{\partial}{\partial r^\alpha_i}\left[\delta^{1/2}_{ij}\left(t^\beta_{ij} + \tilde{t} t^\beta_{ij} - {\tilde t}^2t^\beta_{ij} \right)\right]
  \nonumber \\
  &=& -\frac{1}{2}\delta^{-1}_{ij}\frac{r^\alpha_{ij}}{r_{ij}}{F^t_{ij}}^\beta - k_t\delta^{1/2}_{ij} \left[ (1+{\tilde t}-{\tilde t}^2)\frac{\partial t^\beta_{ij}}{\partial r^\alpha_i} + ({\tilde t}^\beta - 2{\tilde t}{\tilde t}^\beta)\frac{\partial t_{ij}}{\partial r^\alpha_i} + (-{\tilde t}{\tilde t}^\beta+2{\tilde t}^2{\tilde t}^\beta)\frac{\partial t^*_{ij}}{\partial r^\alpha_i} \right]
  \label{Eq:dFsbdra}
\end{eqnarray}
Here we use the notation $\tilde t$ to represent the ratio $t_{ij}/t^*_{ij}$, and the notation ${\tilde t}^\beta$ for ${t_{ij}}^\beta/t^*_{ij}$. The expressions for all the three partial differentiation in~(\ref{Eq:dFsbdra}) are already shown in~(\ref{Eq:dsbdtheta}), (\ref{Eq:dsmagdra}), and (\ref{Eq:ds*dra}).

Similarly, the derivative of tangential force with respect to $\theta_{i}$ (using the same notation as above) can be found as
\begin{equation}
  \frac{\partial {F^t_{ij}}^\beta}{\partial \theta_i} = -k_t\delta^{1/2}_{ij} \left[ (1+{\tilde t}-{\tilde t}^2)\frac{\partial t^\beta_{ij}}{\partial \theta_i} + ({\tilde t}^\beta - 2{\tilde t}{\tilde t}^\beta)\frac{\partial t_{ij}}{\partial \theta_i} \right]
  \label{Eq:dFsbdtheta}
\end{equation}
From the above two equations it is then understood that if ${\B r}_{ij}$ and ${\B t}_{ij}$ are known the differential equations can be solved easily. When ${\tilde t}^\beta$ is negligible for all $\beta$, then ${\tilde t}\approx 0$. This translates to $\frac{\partial {F^t_{ij}}^\beta}{\partial \theta_i} = -(-1)^\beta k_t\sigma_i\delta^{1/2}_{ij}\frac{r_{ij}^\alpha}{r_{ij}}$ with $\alpha \neq \beta$, implying that even in the case of zero tangential displacement and therefore, zero tangential force, the above derivative can be finite.

\subsection{Derivative of normal force}
The derivative of normal force (equation~(\ref{Eq:Fn})) with respect to $r^\alpha_{i}$:
\begin{eqnarray}
  \frac{\partial {F^n_{ij}}^\beta}{\partial r^\alpha_i} &=& k_n \frac{\partial }{\partial r^\alpha_i} \left[ \delta^{3/2}_{ij} \frac{r^\beta_{ij}}{r_{ij}} \right]
  \nonumber \\
  &=& k_n \delta^{1/2}_{ij}\left[ \Delta_{\alpha\beta}\frac{\delta_{ij}}{r_{ij}} - \frac{3}{2}\frac{r^\alpha_{ij} r^\beta_{ij}}{r^2_{ij}} -\left(\frac{\delta_{ij}}{r_{ij}}\right) \frac{r^\alpha_{ij} r^\beta_{ij}}{r^2_{ij}} \right],
  \label{Eq:dFnbdra}
\end{eqnarray}
where $\Delta_{\alpha\beta}$ is the Kronecker delta. The derivative of total force which reads:
\begin{eqnarray}
\label{Eq:dFbdra}
\frac{\partial {F_{ij}}^\beta}{\partial r^\alpha_i} &=& \frac{\partial {F^n_{ij}}^\beta}{\partial r^\alpha_i} + \frac{\partial {F^t_{ij}}^\beta}{\partial r^\alpha_i} \\
\frac{\partial {F_{ij}}^\beta}{\partial \theta_i}  &=& \frac{\partial {F^t_{ij}}^\beta}{\partial \theta_i}
\label{Eq:dFbdtheta}
\end{eqnarray}
can be solved using~(\ref{Eq:dFnbdra}),~(\ref{Eq:dFsbdra}), and~(\ref{Eq:dFsbdtheta}).

\subsection{Derivative of Torque}
The torque of particle-$j$ due to tangential force ${\B F^t}_{ij}$ is ${\B T}_j = -\sigma_j\left({\hat r}_{ij} \times {\B F^t}_{ij}\right) \equiv \sigma_j \tilde{\B{T}}_{ij}$.
In 2D, $\tilde{\B T}_{ij}$ has only z-component:
\begin{equation}
  {\tilde T}_{ij}^z = - \left[ \left(\frac{x_{ij}}{r_{ij}}\right){F^t_{ij}}^y - \left(\frac{y_{ij}}{r_{ij}}\right){F^t_{ij}}^x\right].
  \label{Eq:Tz}
\end{equation}
The derivative of ${\tilde T}_{ij}^z$ then becomes:
\begin{equation}
  \label{Eq:dTdra}
  \frac{\partial {\tilde T}_{ij}^z}{\partial r^\alpha_i} = -\left(\frac{\delta_{\alpha x}}{r_{ij}} - \frac{x_{ij}r_{ij}^\alpha}{r_{ij}^3}\right){F^t_{ij}}^y -  \left(\frac{x_{ij}}{r_{ij}}\right)\frac{\partial {F^t_{ij}}^y}{\partial r^\alpha_i} + \left(\frac{\delta_{\alpha y}}{r_{ij}} - \frac{y_{ij}r_{ij}^\alpha}{r_{ij}^3}\right){F^t_{ij}}^x +  \left(\frac{y_{ij}}{r_{ij}}\right)\frac{\partial {F^t_{ij}}^x}{\partial r^\alpha_i},
\end{equation}
where $\delta_{\alpha x}$ (similarly,  $\delta_{\alpha y}$) is the Kronecker delta, such that $\delta_{x x}=1$ and $\delta_{y x}=0$, and
\begin{equation}
  \frac{\partial {\tilde T}_{ij}^z}{\partial \theta_i} =  - \left[ \left(\frac{x_{ij}}{r_{ij}}\right)\frac{\partial {F^t_{ij}}^y}{\partial \theta_i} - \left(\frac{y_{ij}}{r_{ij}}\right)\frac{\partial {F^t_{ij}}^x}{\partial \theta_i}\right]
  \label{Eq:dTdtheta}
\end{equation}
The above two differential equations can be solved using~(\ref{Eq:dFsbdra}), and~(\ref{Eq:dFsbdtheta}). If the tangential displacement $t_{ij}^\beta$ is negligible compared to the threshold $t_{ij}^*$, i.e., ${\tilde t}^\beta \approx 0$ for all $\beta$. This results in ${\tilde t}\approx 0$. Therefore, $\frac{\partial {\tilde T}_{ij}^z}{\partial \theta_i} = k_t\sigma_i\delta_{ij}^{1/2}$.

\subsection{Jacobian}
The dimension of Jacobian operator $J$ is force over length. To be consistent with the dimension we redefine the torque $T$ and rotational coordinate $\theta$ as
\begin{equation}
  {\tilde T}_i = \frac{T_i}{\sigma_i}, \ \ \ \textrm{and} \ \ \ {\tilde \theta}_i = \sigma_i\theta_i
  \label{Eq:rescaledT}
\end{equation}
In addition, the dynamic matrix has a contribution from the moment of inertia $I_i=I_0m_i\sigma_i^2$ as $\Delta{\B \omega}_i={\B T}_i/I_i\Delta t$. In our calculation, we assume that mass $m_i$ and $I_0$ both are one. The remaining contribution of $I_i$, i.e. $\sigma_i^2$, is taken care of by rescaling the torque and the angular displacement as ${\tilde T}_i$ and ${\tilde \theta_i}$~(\ref{Eq:rescaledT}). For $I_0\neq 1$, the contribution of $I_0$ can be correctly anticipated if we rewrite~(\ref{Eq:dsdt}) as below:
\begin{equation}
  \frac{\mathrm d {\B t}_{ij}}{\mathrm d t} = {\B v}_{ij} - {\B v}^n_{ij} + \frac{1}{I_0}\hat{r}_{ij}\times(\sigma_i{\B \omega}_i + \sigma_j{\B \omega}_j),
  \label{Eq:dsdtnew}
\end{equation}

$J$ essentially contains {\bf four} different derivatives:
\begin{itemize}
\item First type: Derivative of force with respect to the position of particles:
  \begin{equation}
    \label{Eq:Jabij}
    \begin{split}
      & J^{\alpha\beta}_{ij} = \sum_{k=0; k\neq j}^{N-1}\frac{\partial F^\beta_{kj}}{\partial r^\alpha_{i}} = \frac{\partial F^\beta_{ij}}{\partial r^\alpha_{i}}, \ \ \ \textrm{for}\ \ i\neq j \\
      & J^{\alpha\beta}_{ii} = \sum_{j=0; j\neq i}^{N-1}\frac{\partial F^\beta_{ji}}{\partial r^\alpha_{i}} = -\sum_{j=0; j\neq i}^{N-1} J^{\alpha\beta}_{ij},
    \end{split}
  \end{equation}
where $N$ is the total number of particles. $J^{\alpha\beta}_{ij}$ is symmetric if we change pairs, i.e.: $J^{\alpha\beta}_{ij}=J^{\alpha\beta}_{ji}$, however the symmetry is not guaranteed with the interchange of $\alpha$ and $\beta$.
\item Second type: Derivative of force with respect to rotational coordinate:
   \begin{equation}
    \label{Eq:Jbij}
    \begin{split}
      & J^{\beta}_{ij} = -\sum_{k=0; k\neq j}^{N-1}\frac{\partial F^\beta_{kj}}{\partial \tilde\theta_{i}} = -\frac{\partial F^\beta_{ij}}{\partial \tilde\theta_{i}}, \ \ \ \textrm{for}\ \ i\neq j \\
      & J^{\beta}_{ii} = -\sum_{j=0; j\neq i}^{N-1}\frac{\partial F^\beta_{ji}}{\partial \tilde\theta_{i}} = \sum_{j=0; j\neq i}^{N-1} J^{\beta}_{ij}.
    \end{split}
   \end{equation}
   The negative sign makes sure that in stable systems all the eigenvalues are positive. $J^{\beta}_{ij}$ is asymmetric: $J^{\beta}_{ij}=-J^{\beta}_{ji}$.
 \item Third type: Derivative of torque with respect to position:
   \begin{equation}
     \label{Eq:Jaij}
     \begin{split}
       & J^{\alpha}_{ij} = \sum_{k=0; k\neq j}^{N-1}\frac{\partial {\tilde T}_{kj}^z}{\partial r^\alpha_{i}} = \frac{\partial {\tilde T}_j}{\partial r^\alpha_{i}}, \ \ \ \textrm{for}\ \ i\neq j \\
       & J^{\alpha}_{ii} = \sum_{j=0; j\neq i}^{N-1}\frac{\partial {\tilde T}_{ji}^z}{\partial r^\alpha_{i}} = \sum_{j=0; j\neq i}^{N-1} J^{\alpha}_{ij}.
     \end{split}
   \end{equation}
$J^{\alpha}_{ij}$ is also asymmetric: $J^{\alpha}_{ij}=-J^{\alpha}_{ji}$.
 \item Fourth type: Derivative of torque with respect to rotational coordinate:
   \begin{equation}
     \label{Eq:Jij}
     \begin{split}
       & J_{ij} = -\sum_{k=0; k\neq j}^{N-1}\frac{\partial {\tilde T}_{kj}^z}{\partial \tilde\theta_{i}} = -\frac{\partial {\tilde T}_j}{\partial \tilde\theta_i}, \ \ \ \textrm{for}\ \ i\neq j \\
       & J_{ii} = -\sum_{j=0; j\neq i}^{N-1}\frac{\partial {\tilde T}_{ji}^z}{\partial \tilde\theta_{i}} = -\sum_{j=0; j\neq i}^{N-1} J_{ij}.
     \end{split}
   \end{equation}
   The negative sign makes sure that in stable systems all the eigenvalues are positive. $J_{ij}$ is symmetric: $J_{ij}=J_{ji}$.
\end{itemize}

\subsection{Arrangement of Jacobian matrix}
In two dimension $D=2$, for $N$ particles the total number of elements in $J$ is $(D+1)N\times(D+1)N$. In the matrix, first $DN\times DN$ elements contain the first type of force derivative, i.e. $J^{\alpha\beta}_{ij}$. Here the row-index $ro$ and column-index $co$ of $J$ runs in the range $0 \leq ro < DN$ and $0 \leq co < DN$. Rows from $DN \leq ro < (D+1)N$ and columns $0 \leq co < DN$ of $J$ contain $J^\beta_{ij}$, i.e., the second type of derivative. Rows from $0 \leq ro < DN$ and columns $DN \leq co < (D+1)N$ of $J$ contain the third type $J^\alpha_{ij}$. Finally, rows from $DN \leq ro < (D+1)N$ and columns $DN \leq co < (D+1)N$ of $J$ hold $J_{ij}$, i.e., the fourth type of derivative. For a fixed type of derivative, at a fixed row, the column-index first runs over $j$ starting from 0 to $N-1$. Then $\beta$ is incremented, if it exists for that particular derivative type. Similarly, at a fixed column, row-index first runs over $i \in [0, N)$ and then $\alpha$ is incremented.

\end{document}